%% file: main.tex
\documentclass[sigconf]{acmart}
\AtBeginDocument{%
  }
    
\setcopyright{acmlicensed}
\copyrightyear{2024}
\acmYear{2024}

\acmConference[Conference RecSys '24]{Make sure to enter the correct
  conference title from your rights confirmation email}{June 03--05,
  2018}{Woodstock, NY}
\acmISBN{978-1-4503-XXXX-X/18/06}

\usepackage{microtype}
\usepackage{pgfplotstable}
\usepackage{multirow}
\usepackage{array}
\usepackage{makecell}
\usepackage{xcolor}
\usepackage{color}
\usepackage{subcaption}
\usepackage{colortbl}
\usepackage{xspace}
\usepackage{enumitem}

\newcommand{\modelname}{EmbSum\xspace}
\newcommand{\sos}{\texttt{[SOS]}\xspace}
\newcommand{\eos}{\texttt{[EOS]}\xspace}

\copyrightyear{2024}
\acmYear{2024}
\setcopyright{acmlicensed}\acmConference[RecSys '24]{18th ACM Conference on Recommender Systems}{October 14--18, 2024}{Bari, Italy}
\acmBooktitle{18th ACM Conference on Recommender Systems (RecSys '24), October 14--18, 2024, Bari, Italy}

\begin{document}

\renewcommand{\tableautorefname}{Table}
\renewcommand{\sectionautorefname}{Section}
\renewcommand{\subsectionautorefname}{Section}
\renewcommand{\subsubsectionautorefname}{Section}
\renewcommand{\figureautorefname}{Figure}
\renewcommand{\equationautorefname}{Equation}
\newcommand{\linenoautorefname}{Line}
\newcommand\blfootnote[1]{%
  \begingroup
  \renewcommand\thefootnote{}\footnote{#1}%
  \addtocounter{footnote}{-1}%
  \endgroup
}
\title{EmbSum: Leveraging the Summarization Capabilities of Large Language Models for Content-Based Recommendations}


\author{Chiyu Zhang$^\dagger$}
\authornote{Corresponding Authors: chiyuzh@mail.ubc.ca; sunyifei@meta.com}
\affiliation{%
  \institution{University of British Columbia, Canada}
  \country{}
}

\author{Yifei Sun}
\authornotemark[1]
\affiliation{%
  \institution{Meta AI, USA}
  \country{}
}

\author{Minghao Wu}
\affiliation{%
  \institution{Monash University, Australia}
  \country{}
}

\author{Jun Chen}
\affiliation{%
  \institution{Meta AI, USA}
  \country{}
}

\author{Jie Lei}
\affiliation{%
  \institution{Meta AI, USA}
  \country{}
}

\author{Muhammad Abdul-Mageed}
\affiliation{%
  \institution{University of British Columbia, Canada}
  \country{MBZUAI, UAE}
}

\author{Rong Jin}
\affiliation{%
  \institution{Meta AI, USA}
  \country{}
}

\author{Angli Liu}
\affiliation{%
  \institution{Meta AI, USA}
  \country{}
}

\author{Ji Zhu}
\affiliation{%
  \institution{Meta AI, USA}
  \country{}
}

\author{Sem Park}
\affiliation{%
  \institution{Meta AI, USA}
  \country{}
}

\author{Ning Yao}
\affiliation{%
  \institution{Meta AI, USA}
  \country{}
}

\author{Bo Long}
\affiliation{%
  \institution{Meta AI, USA}
  \country{}
}
\renewcommand{\shortauthors}{Zhang et al.}

\begin{abstract}
Content-based recommendation systems play a crucial role in delivering personalized content to users in the digital world. In this work, we introduce \modelname, a novel framework that enables offline pre-computations of users and candidate items while capturing the interactions within the user engagement history. By utilizing the pretrained encoder-decoder model and poly-attention layers, \modelname derives User Poly-Embedding (UPE) and Content Poly-Embedding (CPE) to calculate relevance scores between users and candidate items. 
\modelname actively learns the long user engagement histories by generating user-interest summary with supervision from large language model (LLM). The effectiveness of \modelname is validated on two datasets from different domains, surpassing state-of-the-art (SoTA) methods with higher accuracy and fewer parameters. Additionally, the model's ability to generate summaries of user interests serves as a valuable by-product, enhancing its usefulness for personalized content recommendations.
\end{abstract}




\maketitle
~\blfootnote{ $^{\dagger}$ {Work done during Meta internship.}}
\input{1.introduction}
\input{2.method}
\input{3.experiments}
\input{4.results}

\input{5.conclusion}

\bibliographystyle{ACM-Reference-Format}
\bibliography{custom}

\appendix
\input{6.appendix}

\end{document}

%% file: 1.introduction.tex
\section{Introduction}\label{sec:intro}

In the thriving digital world, billions of users interact daily with a diverse range of digital content, encompassing news, social media updates, e-books, etc. Content-based recommendation systems, as discussed in various studies \cite{gu-2016-learning, embedding-2017-okura, malkiel-2020-recobert, mao_unitrec_2023}, leverage the textual content, such as news articles and books, and the sequence of a user's interaction history. It facilitates the delivery of content recommendations that are more precise, relevant, and customized to each user's preferences.

Recent studies have successfully incorporated Pretrained Language Models (PLMs) into recommendation systems for processing textual inputs \citep{wu-2021-newsbert,  liu_perconet_2023, liu_once_2023}. This integration has significantly improved the efficiency of content-based recommendations. 
Due to the well-known memory limitation of the attention mechanism, previous studies~\cite{li_miner_2022, zhang_unbert_2021} typically encode each piece of user historical content separately and then aggregate them. This approach, however, falls short in modeling the interactions among the user's historical contents. 
Addressing this, \citet{mao_unitrec_2023} introduce local and global attention mechanisms to encode user histories hierarchically. However, this method needs to truncate the history sequence to 1K tokens due to the limitation of PLMs, thus diminishing the benefits of leveraging extensive engagement history to capture users' comprehensive interests. 
In another vein, to improve the alignment between user and candidate content, many studies have directly integrated candidate items into user modeling~\citep{qi-2022-news, li_miner_2022, xu-2023-candidateaware}. This online real-time strategy prevents recommendation systems from performing offline pre-computations for efficient inference, which restricts their real-world applications.

\begin{figure*}[t]
    \centering
\includegraphics[width=\linewidth]{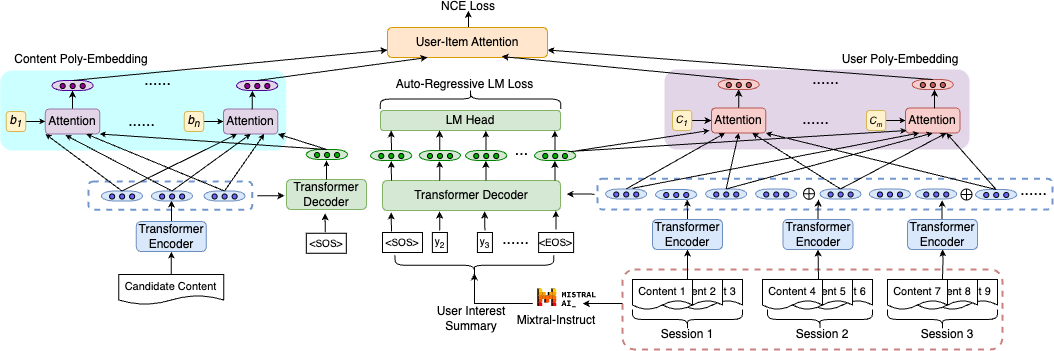}
\caption{Overview of our \modelname framework. Note that the user summaries generated by LLMs are only used in training.}
\label{fig:overview}
\end{figure*}

To tackle the aforementioned challenges, we introduce a new framework, \modelname, which enables offline pre-computations of embeddings of users and candidate items while capturing the interactions within the user's long engagement history. As Figure~\ref{fig:overview} shows, we utilize poly-attention~\cite{humeau_poly-encoders_2020} layers to derive multiple embeddings for the detailed features of both users and candidate items, referred to as User Poly-Embedding (UPE) and Content Poly-Embedding (CPE), respectively. These embeddings are then used to calculate the relevance scores between users and candidate items. More specifically, we use the pretrained T5 encoder \citep{raffel-2020-exploring} to encode user engagement sessions independently. We hypothesize that merely concatenating the embeddings of history sequences does not effectively model the interactions between user engagement sessions. To address this, we use the T5 decoder to fuse session-based encoded sequences by training it to generate user-interest summaries, supervised by the large language models (LLMs)~\cite{jiang-2024-mixtral} generated summaries of user's holistic interests when modeling user representations.

Our contributions in this work can be summarized as follows:
\begin{enumerate}
    \item We present a new framework, \textbf{\modelname}, for embedding and summarizing user interests in content-based recommendation systems. This framework employs an encoder-decoder architecture to encode extensive user engagement histories and produce summaries of user interests.
    \item We validate the effectiveness of \modelname by testing it on two popular datasets from different domains. Our approach surpasses SoTA methods, delivering higher accuracy with fewer parameters.
    \item Our model can generate summaries of user interests, which serves as a beneficial by-product, thereby enhancing its usefulness for personalized content recommendations.
\end{enumerate}

%% file: 2.method.tex
\section{Methodology} \label{sec:method}

In this section, we first describe the problem formulation of our work (\autoref{subsec:problem}). Then, we provide an overview of our \modelname framework (\autoref{subsec:overview}). Next, we introduce the details of modeling user engagements (\autoref{subsec:user_engage}), candidate contents (\autoref{subsec:candidate}), and the click-through predictor and training objectives (\autoref{subsec:train_infer}).

\subsection{Problem Formulation}
\label{subsec:problem}

Given a user $u_i$ and a candidate content item $\bar{e}_j$ (such as news articles or books), the objective is to derive a relevance score $s^i_j$, which indicates the likelihood of user $u_i$ engaging with (e.g., clicking on) the content item $\bar{e}_j$. Considering a set of candidate contents $C = \{\bar{e}_1, \bar{e}_2, \dots, \bar{e}_j\}$, these contents are ranked based on their relevance scores $\{s^i_1, s^i_2, \dots, s^i_j\}$ for user $u_i$.
It is crucial to effectively extract user interests from their engagement history. User $u_i$ is characterized by a sequence of $k$ historically engaged contents $E_{u_i}$ (such as browsed news articles or positively rated books), sorted in descending order by engagement time.

\subsection{Overview of \modelname}
\label{subsec:overview}

\autoref{fig:overview} presents an overview of our proposed model, \modelname. Both the user engagements and the candidate contents are encoded using a pretrained encoder-decoder Transformer model. The click-through rate (CTR) predictions between the users and the candidate content are modeled using noisy contrastive estimation (NCE) loss. Moreover, we also introduce a user-interest summarization objective that is supervised by LLM generations. By leveraging both poly-embedding of user engagement history and candidate content, 
\modelname can identify patterns and preferences from user interactions and match them with the most suitable content.

\subsection{User Engagement Modeling}
\label{subsec:user_engage}

\paragraph{\textbf{Session Encoding}}
The user engagement history, denoted as $E_{u_i}$, comprises a sequence of $k$ content items that a user has previously engaged with. To address the high memory demands of processing long sequences with attention mechanisms, we partition these $k$ content items into $g$ distinct sessions, represented as $E_{u_i} = \{\eta_1, \eta_2, \dots, \eta_g\}$. Each session $\eta_i$ encapsulates $l$ tokens from $p$ content items, expressed as $\eta_i=\{e_1, e_2, \dots, e_p\}$.
This structure is designed to reflect the user's interests over specific time periods and to improve the interactions within each session. We encode each session using a T5 encoder~\cite{raffel-2020-exploring} independently. The representation of each piece of content is then derived from the hidden state output corresponding to its first token, which is the start-of-sentence symbol \sos. This process yields $k$ representation vectors.

\paragraph{\textbf{User Engagement Summarization}}
We posit that merely utilizing these $k$ representations falls short in capturing the subtleties of user interests and the dynamics among long-range engaged contents. To address this, we exploit the capabilities of SoTA LLMs to synthesize a distilled summary of user interests, given that recent studies \citep{DBLP:journals/corr/abs-2303-08774, DBLP:journals/corr/abs-2312-11805} have demonstrated LLMs' capacity for summarizing long sequences. This summary encapsulates a rich and comprehensive perspective of user preferences. We utilize \texttt{Mixtral-8x22B-Instruct}~\cite{jiang-2024-mixtral} to generate these interest summaries from the engagement histories.
The resulting summaries are then incorporated into the T5 decoder. Drawing inspiration from the fusion-in-decoder concept~\cite{izacard_leveraging_2021}, we concatenate the hidden states of all tokens from all subsequences encoded in a session-based manner and input this combined sequence into the T5 decoder. We then train the model to produce a summary of the user's interests, employing the following loss:
\begin{equation}
\label{eq:sum_loss}
\mathcal{L}_{\text{sum}} = - \sum_{j=1}^{|y^{u_i}_j|} log (p(y^{u_i}_j|E, y^{u_i}_{<j})),
\end{equation}
where $y^{u_i}_j$ represents the summary generated for user $u_i$, and $|y^{u_i}_j|$ is the length of the user-interest summary. 

\paragraph{\textbf{User Poly-Embedding}}
Given that each session is encoded independently, a global representation for all engaged contents is also necessary. 
We hence acquire a global representation that is derived from the last token of the decoder output (i.e., \eos token), representing all user engagements collectively. We then concatenate it with the $k$ representation vectors from session encoding to form a matrix $Z\in \mathrm{R}^{(k+1)\times d}$. With these representations of user interaction history, we employ a poly-attention layer~\cite{humeau_poly-encoders_2020} to extract the user's nuanced interests into multiple representations. The computation of each user-interest vector $\alpha_a$ is as follows:
\begin{equation}
 \begin{aligned}
    \alpha_a &= \mathrm{softmax}\left[c_a\mathrm{tanh}(ZW^f)^\top\right] Z,
\end{aligned}\label{eq:polyattention}
\end{equation}
where $c_a \in \mathrm{R}^{1\times p}$ and $W^f\in \mathrm{R}^{d\times p}$ are the trainable parameters. We then concatenate $m$ user-interest vectors into a matrix $A\in \mathrm{R}^{m\times d}$, which serves as the user representation, referring to as the User Poly-Embedding (UPE) in our framework.

\subsection{Candidate Content Modeling}
\label{subsec:candidate}

Unlike the conventional practice of using only the first token of a sequence for representation, we introduce the novel Content Poly-Embedding (CPE). Similar to UPE, this method employs a set of context codes, denoted as $\{b_1, b_2, \dots, b_n\}$, to create multiple embeddings for a piece of candidate content. 
For each piece of candidate content, we generate a corresponding vector $\beta_a$ through a poly-attention layer defined by the equation referenced as \autoref{eq:polyattention}, which includes a trainable parameter $W^o$. The resulting $n$ vectors for candidate content are then aggregated into a matrix $B \in \mathrm{R}^{n \times d}$, providing a more nuanced representation that is expected to improve the performance of relevance scoring in the user-candidate matching predictor.

\subsection{CTR Prediction and Training}
\label{subsec:train_infer}
\paragraph{\textbf{CTR Prediction}}
To compute the relevance score \( s^i_j \), we first establish the matching scores between the user representation embedding \( A_i \) and the candidate content representation embedding \( B_j \). This computation is performed using the inner product, followed by flattening the resultant matrix:
\begin{equation}
    K^i_j = \text{flatten} (A_i^\top B_j). 
\end{equation}
where $ K^i_j \in R^{mn}$ is the flattened attention matrix. Subsequently, an attention mechanism is employed to aggregate the matching scores represented by this flattened vector:
\begin{equation}
\begin{aligned}
    W^p &= \text{softmax}(\text{flatten}(A \cdot \text{gelu}(B W^s)^\top)),\\
    s^i_j &= W^p \cdot K^i_j,
\end{aligned}
\end{equation}
where \( W^s \in R^{d \times d} \) signifies a trainable parameter matrix, \( W^p \in R^{mn} \) represents the attention weights obtained after flattening and the softmax function, and \( s^i_j \) is the scaled relevance score.

\paragraph{\textbf{Training}}
We follow the method of training the end-to-end recommendation models using the NCE loss \citep{wu_empowering_2021,liu_once_2023}:
\begin{equation} 
\mathcal{L}_{\text{NCE}} = -\log \left( \frac{\exp(s^i_+)}{\exp(s^i_+) + \sum_{j} \exp(s^i_{-,j})} \right), 
\end{equation}
where \( s^i_+ \) represents the score of the positive sample with which the user engaged, and \( s^i_- \) represents the scores of negative samples. 
Therefore, the overall loss function is defined as: 
\begin{equation} 
\mathcal{L} = \mathcal{L}_{\text{NCE}} + \lambda \mathcal{L}_{\text{sum}},
\end{equation}
where \( \lambda \) is a scaling factor determined to be 0.05 based on the performance on the validation set.

%% file: 3.experiments.tex
\section{Experiments}

\paragraph{\textbf{Baselines}} 
We evaluate our~\modelname~against a range of commonly used and SoTA neural network-based content recommendation approaches. These include methods that train text encoders from scratch, such as (1) NAML~\cite{wu-2019-neural}, (2) NRMS~\cite{neural-2019-wu}, (3) Fastformer~\cite{wu_fastformer_2021}, (4) CAUM~\cite{qi-2022-news}, and (5) MINS~\cite{wang-2022-news}. We also consider systems that utilize PLMs, including (6) NAML-PLM, (7) UNBERT~\cite{zhang_unbert_2021}, (8) MINER~\cite{li_miner_2022}, and (9) UniTRec~\cite{mao_unitrec_2023}. More Details on these baselines and our implementation are provided in \autoref{append:baselines}.

\paragraph{\textbf{Dataset}}
We employ two publicly available benchmark datasets for content-based recommendation. The first dataset, MIND~\cite{wu_mind_2020}, consists of user engagement logs from Microsoft News, incorporating both positive and negative labels determined by user clicks on news articles. We use the smaller version of this dataset, which includes 94K users and 65K news articles. The second dataset is obtained from Goodreads~\cite{wan-2018-item}, which focuses on book recommendations derived from user ratings. In this dataset, ratings above 3 are considered positive labels, while ratings below 3 are regarded as negative labels. The Goodreads dataset comprises 50K users and 330K books. More details on dataset statistics are at \autoref{sec:data_stat}.

\paragraph{\textbf{Evaluation}} We utilize a variety of metrics to assess the performance of content-based recommendation systems. These metrics include the classification-based metric AUC~\cite{fawcett-2006-pattern}, and ranking-based metrics such as MRR~\cite{voorhees-1999-trec8} and nDCG@top$N$ (with top$N$ = 5, 10)~\cite{jarvelink-2002-cumulated}. Metric calculations are performed using the Python library TorchMetrics~\cite{detlefsen-2022-torchmetrics}. We determine the best model on the Dev set using AUC and report the corresponding Test performance.

\paragraph{\textbf{Content Formatting}}
As an example, for each content in MIND dataset, we combine its fields into a single text sequence using the following template: ``News Title: $\left<title\right>$; News Abstract: $\left<abstract\right>$; News Category: $\left<category\right>$''.

\begin{figure}
    \centering
    \includegraphics[width=\linewidth]{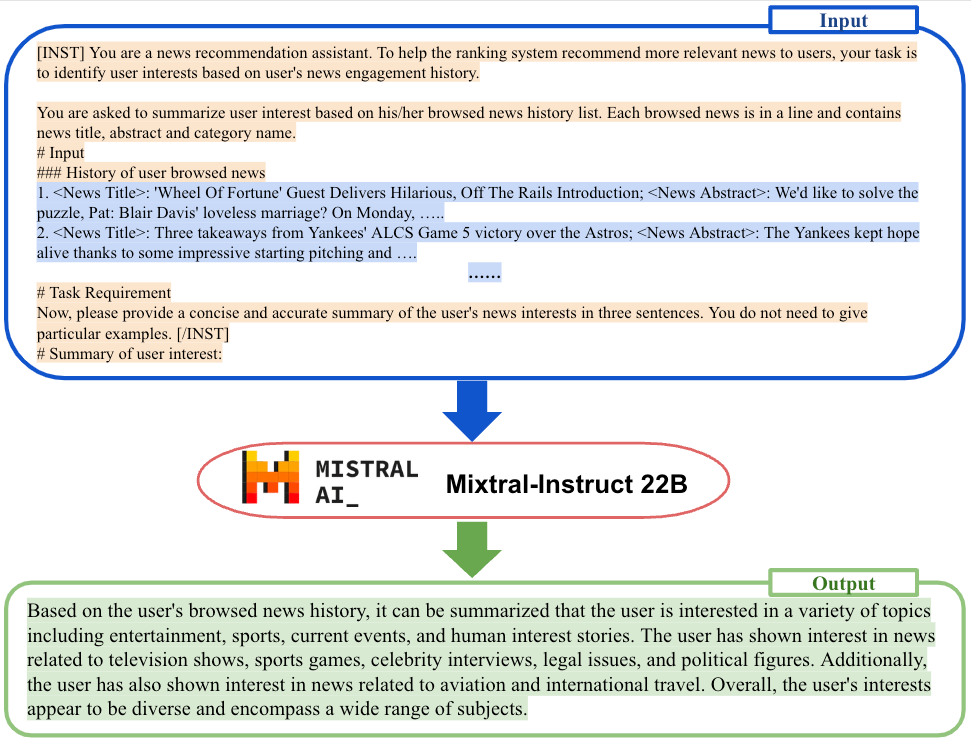}
    \caption{Illustration of using an LLM for user interest profiling. The input provided to the LLM is enclosed in a red box, and the output generated by the LLM is shown in a green box. The segment marked in orange within the input specifies the instruction for the task, whereas the portion in blue highlights the history of news browsed by the user.}
    \label{fig:prompt_llm}
\end{figure}

\paragraph{\textbf{LLM Based User-Interest Summary}}
We leverage the open-source \texttt{Mixtral-8x22B-Instruct}~\cite{jiang-2024-mixtral} to create summaries that reflect users' interests based on their engagement history. Figure~\ref{fig:prompt_llm} illustrates an example of the input provided and the corresponding summary generated for the MIND dataset. The process starts with an instruction to frame the task, followed by a list of news items the user has viewed, ordered from the most recent to the oldest. Each item includes its title, abstract, and category. The input is limited to 60 engagement items, with extended news abstracts or book descriptions being condensed to 100 words. The instruction concludes with a request for the model to condense the user's interests into three sentences. The latter part of Figure~\ref{fig:prompt_llm} shows a sample output from the LLM. Additionally, we evaluate the LLM-generated user-interest summaries using GPT-4 (i.e., \texttt{gpt-4o} API) as a judge. The results indicate that most of the generated summaries accurately capture the user’s interests. Further details of this experiment are provided in Section~\ref{sec:summary_quality} of Appendix. 
As shown in Table~\ref{tab:dataset} in Appendix, the average length of summaries generated by this process is 76 tokens for the MIND dataset and 115 tokens for the Goodreads dataset.

\paragraph{\textbf{Implementation}} 
We utilize the pretrained T5-small model~\cite{raffel-2020-exploring}, which comprises 61M parameters. After hyperparameter tuning, we determined that the optimal codebook sizes for the UPE and CPE layers are 32 and 4, respectively, for both datasets. The model is trained with a learning rate of $5e-4$ and a batch size of 128 over 10 epochs. In all experiments, we cosnsider the latest 60 interactions as the user's engagement history.
For the MIND dataset, we apply a negative sampling ratio of 4, limit news titles to 32 tokens, and restrict news abstracts to 72 tokens. Including approximately 20 additional tokens for the news category and template, each user's engaged history in MIND can total up to 7,440 tokens. In the Goodreads dataset, we set the negative sampling ratio to 2, limit book titles to 24 tokens, and constrain book descriptions to 85 tokens. Consequently, a user in the Goodreads dataset can have up to 7,740 tokens for their engagement history.

%% file: 4.results.tex
\input{table_pic/results_table}
\section{Results}
\paragraph{\textbf{Main Results}}
\autoref{tab:main_res} shows the test results of nine baselines and our \modelname. We can find that models initialized with PLMs obtained much better performance than baselines trained from scratch. 
We observe that \modelname outperforms previous SoTA AUC scores given by UNBERT \citep{zhang_unbert_2021}.
Compared to UNBERT, \modelname achieves an improvement of 0.22 and 0.24 AUC on the MIND and Goodreads datasets, respectively. Notably, \modelname uses only T5-small as the backbone, which has 61M parameters, significantly fewer than the 125M parameters of BERT-based methods (e.g., UNBERT and MINER). On other ranking-based metrics, \modelname achieves the best MRR and nDCG@10 scores on both datasets. \modelname outperforms UniTRec which also uses an encoder-decoder architecture but cannot produce standalone user and candidate embeddings.

\input{table_pic/ablation}
\paragraph{\textbf{Ablation Studies}}
To better understand the effectiveness of our framework, we conduct ablation studies on both datasets, the results of which are presented in \autoref{tab:ablation}. 
We first remove the CPE for the candidate item and use only the encoder hidden state of the \sos token to represent the candidate content. This alteration consistently results in the largest performance drop on both datasets, decreasing the AUC scores by 3.78 and 0.67 on the MIND and Goodreads datasets, respectively. These results suggest the effectiveness of utilizing multiple embeddings to represent candidate content and enhance user-item interactions. Removing session-based grouping and encoding each content separately leads to AUC drops of 0.61 and 0.25 on the MIND and Goodreads datasets, respectively. When we reduce the codebook size of the UPE layer to 1, meaning each user is represented by a single vector, it results in a performance decrease of 0.54 and 0.29 AUC on the MIND and Goodreads datasets, respectively. This also justifies the efficacy of using multiple embeddings in EmbSum. Furthermore, we investigate the efficacy of using LLM-generated user-interest summaries by removing $\mathcal{L}_{sum}$. Without $\mathcal{L}_{sum}$, we only provide the \sos token as the decoder input and take the hidden state of the \sos token as the global representation. As Table~\ref{tab:ablation} shows, this change results in a performance drop on both datasets (0.52 on MIND and 0.14 on Goodreads).

\begin{figure}[t]
    \centering
\includegraphics[width=\linewidth]{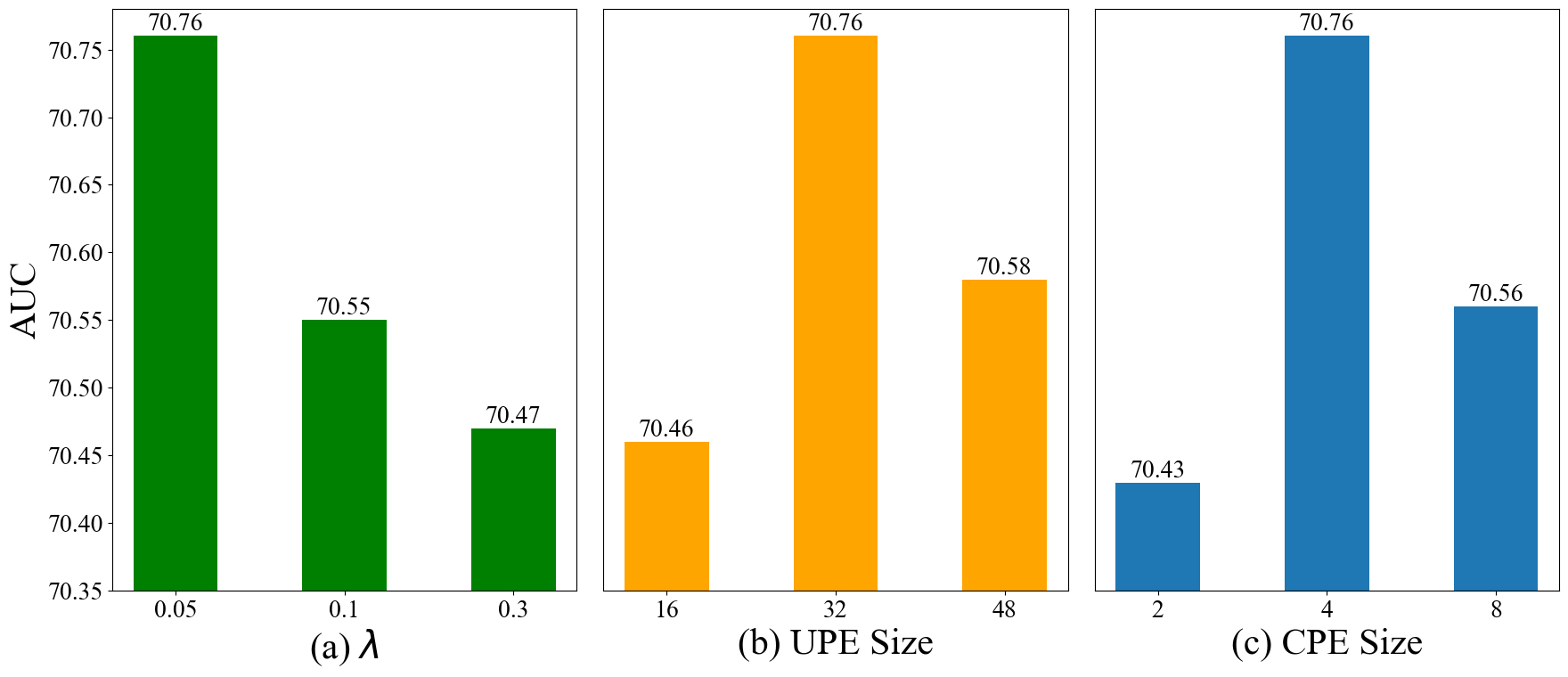}
\caption{Influence of different hyperparameters.}
\label{fig:hyperparmeters}
\end{figure}

\paragraph{\textbf{Influence of Hyperparameters.}} We investigate the model sensitivity to the weight for summarization loss ($\lambda$) and the size of CPE and UPE. We randomly sampled 20\% Train data of MIND dataset for training and validated performance on the Dev set. As Figure~\ref{fig:hyperparmeters}a shows, the ($\lambda$) values of 0.05, 0.1, and 0.3 yielded Dev performances of 70.76, 70.55, and 70.47, respectively. The CPE size values of 2, 4, and 8 resulted in Dev performances of 70.43, 70.76, and 70.56, respectively. The UPE size values of 16, 32, and 48 achieved Dev performances of 70.46, 70.76, and 70.58, respectively. These results indicate that our EmbSum model is not highly sensitive to hyperparameters. For CPE size, our findings suggest that increasing the number of context codes on the candidate item side does not improve model performance. We believe that excessively numerous codebooks for a single item might introduce superfluous parameters, thereby negatively impacting performance. 

\input{table_pic/summary_result}
\paragraph{\textbf{\modelname-Generated Summary}} 
We also evaluate the quality of the summaries generated by our model and report the ROUGE scores in \autoref{tab:summary_eval}. Since CTR is our main task, we generate summaries using the checkpoint that achieves the best AUC score on the Dev set. We select the test users who are not included in the training set and then generate their interest summaries based on their engagement history. We use the summaries generated by \texttt{Mixtral-8x22B-Instruct} as references for calculating ROUGE scores. For the MIND and Goodreads datasets, our model achieves ROUGE-L scores of 39.12 and 28.16, respectively. More concrete examples of the \modelname-generated summaries can be found at \autoref{sec:sum_example}, demonstrating that \modelname is capable of accurately capturing the user's diverse interests.

%% file: table_pic/results_table.tex
\begin{table}[t]
\centering
\setlength\tabcolsep{6pt}
\begin{tabular}{@{}lcccc@{}}
\toprule
\multicolumn{5}{c}{\textbf{MIND}}                                           \\ \midrule
              & AUC            & MRR            & nDCG@5         & nDCG@10        \\ \midrule
NAML          & 66.10          & 34.65          & 32.80          & 39.14          \\
NRMS          & 63.28          & 33.10          & 31.50          & 37.68          \\
Fastformer    & 66.32          & 34.75          & 33.03          & 39.30          \\
CAUM          & 62.56          & 34.40          & 32.88          & 38.90          \\
MINS          & 61.43          & 35.99          & 34.13          & 40.54          \\\hline
NAML-PLM      & 67.01          & 35.67          & 34.10          & 40.32          \\
UNBERT        & \underline{71.73}          & 38.06          & \underline{36.67}          & \underline{42.92}          \\
MINER         & 70.20          & \underline{38.10}          & 36.35          & 42.63          \\
UniTRec       & 69.38          & 37.62          & 36.01          & 42.20          \\\hline
EmbSum (ours) & \textbf{71.95} & \textbf{38.58} & \textbf{36.75} & \textbf{42.97} \\ \midrule
\multicolumn{5}{c}{\textbf{Goodreads}}                                             \\ \midrule
NAML          & 59.35          & 72.16          & 53.49          & 67.81          \\
NRMS          & 60.51          & 72.15          & 53.69          & 68.03          \\
Fastformer    & 59.39          & 71.11          & 52.38          & 67.05          \\
CAUM          & 55.13          & 73.06          & \textbf{54.97} & \underline{69.02}          \\
MINS          & 53.02          & 71.81          & 53.72          & 68.00          \\\hline
NAML-PLM      & 59.57          & 72.54          & 53.98          & 68.41          \\
UNBERT        & \underline{61.40}          & \underline{73.34}          & 54.67          & 68.71          \\
MINER         & 60.72          & 72.72          & 54.17          & 68.42          \\
UniTRec       & 60.00          & 72.60          & 53.73          & 67.96          \\\hline
EmbSum (ours) & \textbf{61.64} & \textbf{73.75} & \underline{54.86}          & \textbf{69.08} \\ \bottomrule
\end{tabular}
\caption{Results on MIND-small and Goodreads. The best results are highlighted in \textbf{bold}. The second-best results are highlighted in \underline{underscore}.}\label{tab:main_res}
\end{table}

%% file: table_pic/ablation.tex
\begin{table}[t]
\centering
\begin{tabular}{@{}lcccc@{}}
\toprule
\multicolumn{5}{c}{\textbf{MIND}}               \\ \midrule
                   & AUC   & MRR   & nDCG@5 & nDCG@10 \\\midrule
Ours               & \textbf{71.95} & 38.58 & \textbf{36.75}  & \textbf{42.97}   \\ \hline
wo CPE             & 68.17 & 36.49 & 33.72  & 40.20   \\
wo grouping        & 71.34 & 38.29 & 36.41  & 42.55   \\
wo UPE             & 71.41 & \textbf{38.64} & 36.70  & 42.90   \\
wo $\mathcal{L}_{sum}$ & 71.43 & 38.42 & 36.39  & 42.60   \\ \midrule
\multicolumn{5}{c}{\textbf{Goodreads}}                \\ \midrule
Ours               & \textbf{61.64} & \textbf{73.75} & \textbf{54.86}  & \textbf{69.08}   \\\hline
wo CPE             & 60.97 & 72.94 & 54.39  & 68.53   \\
wo grouping        & 61.39 & 73.53 & 54.79  & 68.86   \\
wo UPE             & 61.35 & 73.55 & 54.67  & 68.81   \\
wo $\mathcal{L}_{sum}$ & 61.50 & 73.55 & 54.74  & 68.91   \\ \bottomrule
\end{tabular}
\caption{Result of ablation study for \modelname. The best results are highlighted in \textbf{bold}.}
\label{tab:ablation}
\end{table}

%% file: table_pic/summary_result.tex

\begin{table}[t]
\centering
\begin{tabular}{@{}lccc@{}}
\toprule
          & \textbf{ROUGE 1} & \textbf{ROUGE 2} & \textbf{ROUGE L} \\ \midrule
MIND      & 51.42            & 30.33            & 39.12            \\ 
Goodreads & 46.99            & 20.86            & 28.16            \\ \bottomrule
\end{tabular}
\caption{Evaluation on \modelname-generated user summary.}
\label{tab:summary_eval}
\end{table}

%% file: 5.conclusion.tex
\section{Conclusion}\label{sec:conclude}
We present a novel framework~\modelname~ for content-based recommendation. \modelname~utilize encoder-decoder architecture and poly-attention modules to learn independent user and candidate content embeddings as well as generate user-interest summaries based on long user engagement histories. Through our experiments on two benchmark datasets, we have demonstrated that our framework achieves SoTA performance while using fewer parameters, and being able to generate a user-interest summary which can be used for recommendation explainability/transparency.

%% file: 6.appendix.tex
\newpage
\noindent\textbf{\Huge Appendices}

\section{Baselines}\label{append:baselines}

Our model is evaluated in comparison with the prior state-of-the-art neural network-based methods for content-based recommendation:

\begin{enumerate}
\item NAML~\cite{wu-2019-neural} employs a content representation approach that integrates CNNs, additive attention, and pretrained word embeddings. It further uses additive attention to create a user representation based on their engagement history.
\item NRMS~\cite{neural-2019-wu} combines pretrained word embeddings with multi-head self-attention and additive attention mechanisms to develop representations for user preferences and candidate content.
\item Fastformer~\cite{wu_fastformer_2021} presents an efficient Transformer model that consists of additive attention.
\item CAUM~\cite{qi-2022-news} extends NRMS by incorporating content title entities into embeddings and utilizing candidate-aware self-attention for generating user embeddings.
\item MINS~\cite{wang-2022-news} improves upon NRMS with a multi-channel GRU-based network designed to better capture the sequential dynamics in user engagement history.
\item NAML-PLM utilizes a PLM as the content encoder, leveraging a pretrained model rather than training from scratch. RoBERTa-base \citep{liu-2019-roberta} (with 125M parameters) is used as the text encoder.
\item UNBERT~\cite{zhang_unbert_2021} employs a PLM for content encoding, deriving user-content matching indicators at both item and word levels, with RoBERTa-base \citep{liu-2019-roberta} (with 125M parameters) as the backbone model.
\item MINER~\cite{li_miner_2022} uses a PLM for text encoding and introduces a poly-attention mechanism to extract diverse user interest vectors for enhanced user representation, standing out as a leading model on the MIND dataset leaderboard.\footnote{\url{https://msnews.github.io/}} RoBERTa-base \citep{liu-2019-roberta} (with 125M parameters) is the text encoder for MINER.
\item UniTRec~\cite{mao_unitrec_2023} applies an encoder-decoder structure (i.e., BART) to separately encode user history and candidate content, assessing candidate content using perplexity scores from the decoder and a discriminative scoring head.\footnote{Metrics are calculated using predictions from the discriminative scoring head.} BART-base \citep{lewis-etal-2020-bart} (with 139M parameters) is the backbone in our implementation.
\end{enumerate}

We utilize the optimal hyperparameters recommended for these baselines and perform training and evaluations on our dataset splits. For NAML, NRMS, Fastformer, and NAML-PLM, we employ the implementations provided by~\citet{legommenders-2023-liu}. The CAUM and MINS implementations are obtained from \citet{iana_newsreclib_2023}. For UNBERT, MINER, and UniTRec, we use the original scripts released by their respective authors.\footnote{UNBERT: \url{https://github.com/reczoo/RecZoo/tree/main/pretraining/news/UNBERT}, MINER: \url{https://github.com/duynguyen-0203/miner}, UniTRec: \url{https://github.com/Veason-silverbullet/UniTRec}.}

\input{table_pic/data_table}
\section{Dataset Statistics}
\label{sec:data_stat}

In this work, we evaluate our model \modelname on two public benchmark datasets for content-based recommendation. The dataset statistics is presented in \autoref{tab:dataset}.

\section{Quality of LLM Generated User-Interest Summary}\label{sec:summary_quality}

 Due to the lack of gold labels for evaluating generated user-interest summarization and the difficulty of human evaluation, we follow recent works that use LLM as a judge to evaluate AI response. We adapt the rubric from \cite{selfinstruct-2023-wang},\footnote{\url{https://github.com/yizhongw/self-instruct}} which scores the AI response in four levels, A (accurate and comprehensive), B (acceptable but with few minor errors), C (related but has significant errors), and D (invalid and unrelated response). We prompt the SoTA LLM GPT-4 (i.e., \texttt{gpt-4o} API) to evaluate the generated user-interest summaries. To be budget-friendly, we randomly sample 500 users for this evaluation. We provide the original user engagement history and our generated user-interest summary. The GPT-4 judge classifies 187 summaries to A, 305 to B, 8 to C, and 0 to D. We also conducted this evaluation on the generated user-interest summaries of Goodreads users. The GPT-4 judge classifies 156, 331, 13, and 0 summaries to A, B, C and D, respectively. This result indicates that most of our generated summaries are able to capture the user's interests.

\input{table_pic/example}
\section{EmbSum-Generated Summary Examples}
\label{sec:sum_example}
We present the examples of EmbSum-generated summaries in \autoref{tab:sum_example}.

%% file: table_pic/data_table.tex

\begin{table*}[t]
\centering
\begin{tabular}{@{}lrrr|rrr|lrr@{}}
\toprule
     \multicolumn{1}{c}{Dataset}        & \multicolumn{3}{c}{MIND}                                                       & \multicolumn{3}{c|}{Goodreads}                                                  & \multicolumn{1}{c}{\multirow{2}{*}{Dataset}}       & \multicolumn{1}{c}{\multirow{2}{*}{MIND}} & \multicolumn{1}{c}{\multirow{2}{*}{Goodreads}} \\ \cmidrule(lr){1-1} \cmidrule(lr){2-4}\cmidrule(lr){5-7}
   \multicolumn{1}{c}{Split}   & \multicolumn{1}{c}{Train} & \multicolumn{1}{c}{Dev} & \multicolumn{1}{c}{Test} & \multicolumn{1}{c}{Train} & \multicolumn{1}{c}{Dev} & \multicolumn{1}{c|}{Test} & \multicolumn{1}{c}{}       & \multicolumn{1}{c}{}                      & \multicolumn{1}{c}{}                           \\ \midrule
\# content   & 51,283                    & 21,352                  & 41,496                    & 309,047                   & 234,232                 & 247,242                   & \# of history/user         & 22                                        & 47                                             \\
\# users     & 50,000                    & 6,679                   & 46,549                    & 21,450                    & 16,339                  & 17,967                    & \# category                & 18                                        & 11                                             \\
\# new users & \multicolumn{1}{c}{-}     & 5,862                   & 41,020                    & \multicolumn{1}{c}{-}     & 2,930                   & 3,199                     & \# tokens/title            & 17                                        & 18                                             \\
\# positive  & 236,344                   & 10,775                  & 100,608                   & 198,403                   & 75,445                  & 93,156                    & \# tokens/abstract         & 50                                       & 128                                            \\
\# negative  & 5,607,100                 & 249,607                 & 2,380,008                 & 458,435                   & 141,977                 & 154,016                   & \# tokens/user summary & 76                                        & 115                                            \\ \bottomrule
\end{tabular}
\caption{Dataset Statistics. ``\# new users'' indicates the number of users not included in the Train set. ``\# tokens/user summary'' represents the average length of user interest summaries generated by LLM. The number of tokens are calculated using the RoBERTa-base model's vocabulary.}
\label{tab:dataset}
\end{table*}


%% file: table_pic/example.tex
\begin{table*}[t]
\centering
\small
\setlength\tabcolsep{1pt}
\begin{tabular}{cp{0.75\linewidth}c}
\toprule
\multicolumn{3}{c}{\textbf{Historical Clicked News}}                                                                                                                                                                                                                                                                                                           \\ \midrule
(1)                                                                      & 45 Amazing Facts About Airplanes That Will Make Your Mind Soar                                                                                                                                   & travel                                                                            \\
(2)                                                                       & 29 Foods Diabetics Should Avoid                                                                                                                                                                  & health                                                                            \\
(3)                                                                       & This \$12 million 'mansion yacht' is made entirely of stainless steel and it's a first for the industry. Take a peek inside.                                                                      & travel                                                                            \\
(4)                                                                       & The \#1 Worst Menu Option at 76 Popular Restaurants                                                                                                                                               & health                                                                            \\
(5)                                                                       & Celebs celebrate Halloween 2019                                                                                                                                                                  & entertainment                                                                     \\
(6)                                                                       & Woman who made it on Delta flight without a ticket or boarding pass says 'it's not my fault'                                                                                                     & travel                                                                            \\
(7)                                                                       & This Dog's Terrifyingly Cute 'Killer' Costume Just Won Halloween                                                                                                                                 & lifestyle                                                                         \\
(8)                                                                       & Body of missing Alabama girl found; 2 being charged                                                                                                                                              & news                                                                              \\
(9)                                                                       & Meghan Markle Always Stands the Exact Same Way at Events and There's a Specific Reason Why                                                                                                       & lifestyle                                                                         \\
(10)                                                                      & Prince William and Kate Middleton arrive in Pakistan for royal tour                                                                                                                              & lifestyle                                                                         \\ \midrule
\multicolumn{3}{c}{\textbf{EmbSum-Generated Summary}}                                                                                                                                                                                                                                                                                                          \\ \midrule
\multicolumn{3}{p{\linewidth}}{Based on the user's browsed news history, their interests seem to be focused on entertainment, lifestyle, and current events. They appear to enjoy reading about celebrities, royal families, and unusual or quirky stories. Additionally, they seem to have an interest in health and wellness, specifically for weight loss and fitness.} \\ \bottomrule
\end{tabular}
\caption{Example of \modelname-generated summary.}
\label{tab:sum_example}
\end{table*}